\documentclass[12pt]{article}

\usepackage{authblk}
\usepackage{dcolumn}
\usepackage{amsbsy}
\usepackage{amsmath}
\usepackage{fullpage}

\usepackage{amssymb}
\usepackage{color}
\usepackage{graphicx}
\usepackage{epstopdf}
\usepackage{caption}
\usepackage{subcaption}
\usepackage{bm}
\usepackage{caption}
\usepackage{cite}
\usepackage{mathrsfs}
\usepackage{arydshln}
\usepackage{multicol,multirow}
\usepackage{comment}
\usepackage[dvipsnames]{xcolor}

\newcommand{\ba}{\bm{a}}

\newcommand{\bc}{\bm{c}}

\newcommand{\bu}{\bm{u}}

\newcommand{\bI}{\bm{I}}
\newcommand{\bK}{\bm{K}}
\newcommand{\bhK}{\bm{\hat{K}}}

\newcommand{\bM}{\bm{M}}

\newcommand{\bzero}{\mathbf{0}}

\begin{document}
	
\title{A Bloch-based procedure for dispersion analysis of lattices with periodic time-varying properties}
\author[1,2]{Javier Vila\thanks{javier.vila@aerospace.gatech.edu}}
\author[2]{Raj Kumar Pal}
\author[2,3]{Massimo Ruzzene}
\author[2]{Giuseppe Trainiti}
\affil[1]{Department of Continuum Mechanics and Theory of Structures, Universidad Carlos III de Madrid}
\affil[2]{School of Aerospace Engineering, Georgia Institute of Technology}
\affil[3]{School of Mechanical Engineering, Georgia Institute of Technology}

\maketitle


\begin{abstract}
	We present a procedure for the systematic estimation of the dispersion properties of linear discrete systems with periodic time-varying coefficients. The approach relies on the analysis of a single unit cell, making use of Bloch theorem along with the application of a harmonic balance methodology over an imposed solution ansatz. The solution of the resulting eigenvalue problem is followed by a procedure that selects the eigen-solutions corresponding to the ansatz, which is a plane wave defined by a frequency-wavenumber pair. Examples on spring-mass superlattices demonstrate the effectiveness of the method at predicting the dispersion behavior of linear elastic media. The matrix formulation of the problem suggests the broad applicability of the proposed technique. Furthermore, it is shown how dispersion can inform about the dynamic behavior of time-modulated finite lattices. The technique can be extended to multiple areas of physics, such as acoustic, elastic and electromagnetic systems, where periodic time-varying material properties may be used to obtain non-reciprocal wave propagation.   
\end{abstract}

\section{Introduction}
Periodic structures are finding increasing applications in diverse fields of science and engineering due to their unique dynamic properties~\cite{hussein2014dynamics}. 
One growing area of research is controlling the direction of wave propagation using periodic arrays of active materials or meta-materials.
During the last decades, a number of researchers have achieved unidirectional wave propagation by spatio-temporal modulation of material properties. For example, Oliner and Cassedy~\cite{cassedy1963dispersion} investigated  modulated electric circuits and obtained expressions for dispersion curves using continued fractions.  
Using their work, other authors~\cite{felsen1970wave,fante1971transmission,elachi1972electromagnetic} studied time-varying electromagnetic media with the objective of investigating wave dispersion.
Similarly, in mechanical systems, Lurie~\cite{lurie1997effective} studied the dynamics of time varying composites and derived their effective properties.  
Extending this work, Shui et al.~\cite{shui2015novel,shui2014one} recently demonstrated asymmetrical wave  properties in spatiotemporal composites. Furthermore, Wright and Cobbold~\cite{wright2009acoustic,wright2010two} extended multiple scattering theory to phononic crystals with time-varying properties.
Using this concept, Engheta and coworkers~\cite{zanjani2015nems,zanjani2014one} demonstrated 
one-way phonon transport in waveguides by a spatio-temporal  
modulation in graphene. Another example in this line of work is that of Li et al.~\cite{li2014wave}, whereby numerical simulations show tunable wave propagation in helicoidal 
phononic crystals with coupling between in-plane torsional and out-of-plane longitudinal waves. By modulating the 
out-of-plane stiffness in space and  time using external control, they alter the in-plane wave propagation. 
Similarly,  Wang et al.~\cite{wang2015acoustic} investigated both numerically and experimentally a tunable acoustic unidirectional device that provide asymmetrical transmission properties.  

Expoliting this periodic modulation of properties, numerous works have demonstrated non-reciprocal wave propagation. 
For example, 
Fleury et al.~\cite{fleury2014sound} achieved one way transport in acoustic channels using a circulator-like device.  
In a similar line of work, Al{\`u} and coworkers~\cite{estep2014magnetic} demonstrated nonreciprocal excitation in electric circuits by time varying capacitance: a small modulation causes the  counter-propagating modes to have distinct frequencies, which results in 
time-reversal symmetry breaking. Asymptotic analysis is employed to predict frequency splitting for a $3$ port system. Deymier and coworkers~\cite{swinteck2015bulk} recently demonstrated unidirectional wave propagation numerically, 
modulating the spring stiffness periodically in space and time in a one dimensional spring-mass chain. 
The authors obtained asymmetrical dispersion diagrams by extracting the data from extensive numerical simulations using the spectral energy density (SED) method~\cite{thomas2010predicting}.

Most of the studies above rely on numerical simulations, which are time consuming, particularly for higher dimensional structures and are
not convenient for design and optimization studies.  A systematic procedure for obtaining  dispersion information of lattices with periodic time-varying properties is therefore of interest for design and analysis of time-varying systems.
In this paper, we present a technique for obtaining the dispersion diagrams of discretized linear elastic media 
with space-time stiffness variability. The method consists of four steps: the appropriate selection of an extended unit cell, a suitable choice of the solution ansatz, the application of Bloch wave analysis and a filtering of the obtained eigen-solutions 
to identify the frequencies associated with the fundamental plane wave solution. 

In periodic time-varying lattices, the size of a unit cell is based not only on the geometry and time independent material 
properties, but also needs to takes into account the size of the stiffness variability within the system \cite{hussein2014dynamics,brillouin2003wave,kittel2005introduction}. In order to make use of a suitable ansatz, we draw upon the 
Hill-determinant method~\cite{bolotin1962dynamic}, which is widely used to obtain criteria for the stability regions of linear systems with time varying coefficients. Through superposition principle, the motion of a particle is expressed by a plane wave whose amplitude is periodic. Then, we employ a Bloch wave based analysis over the extended unit cell to obtain the eigenvalues for each wavenumber. 
These three steps were followed by Trainiti and Ruzzene~\cite{Trainiti2016} in the application of this analytical method to the analysis of continuous mechanical systems with periodic time-varying coefficients. However, the multiplicity of frequencies in the Hill-determinant method causes eigenvalues whose associated eigenvectors represent a different frequency in displacements to arise, leading to non-physical 
curves in their dispersion diagram. Our final step applies a filtering procedure using the eigenvectors 
such that the dispersion diagram reflects only the possible wavenumber-frequency couples.

The outline of this paper is as follows: Section~\ref{TheorySection} presents the developed procedure for computing the dispersion diagram for spring-mass chains with periodic time varying stiffness. Section~\ref{Sec.3} illustrates its implementation in the analysis of three distinct systems, and the results are validated with numerical solutions. Sec.~\ref{Sec.4} shows the usefulness of dispersion diagrams in predicting the broken time-reversal behavior of infinite and finite systems with time varying coefficients. Finally, Sec.~\ref{ConcSection} summarizes our work and presents potential future research directions.

\section{Theoretical background: }\label{TheorySection}

\subsection{Time varying periodic lattice configuration}\label{InfiniteChainsSection}
We consider a one-dimensional (1D) structure that is characterized by periodically modulated properties. The structure is governed by linear interactions, defined by stiffness coefficients that vary in time, and by constant inertia coefficients. The stiffness modulation is expressed as a traveling wave propagating with veloctiy $v_m = \lambda_m / T_m$, where $\lambda_m$ and $T_m$ respectively denote the spatial wavelength and temporal period of the modulation. Thus, at any given instant of time, it is possible to describe the structure as the assembly of unit cells that are identified by one spatial modulation period $\lambda_m$. Based on this description, and the assumption that the structure has been discretized through a Finite Element procedure, we may express the generalized equation of motion for the $n$-th cell of the assembly as
\begin{equation}\label{EqMotionM}
\bM \ddot{\bu}_{n} + \bK^{(l)}(t) \bu_{n-1} + \bK(t) \bu_{n} + \bK^{(r)}(t) \bu_{n+1} = \bzero.
\end{equation}
where $\bM, \bK$ and $\bu_n$ denote the mass, stiffness matrices and a vector of degrees of freedom of the unit cell $n$, while $ \bK^{(l)}, \bK^{(r)}$ describe the interactions of this unit cell with its neighboring unit cells. For simplicity of notation, but without loss of generality, it is assumed that the mass is lumped so that there is no inertial coupling across the cells.

The stiffness matrices are all expressed as a periodic functions of time with period $T_m$ 
so that the following relation holds
\begin{equation}\label{PeriodicityTime}
\bK(t) = \bK(t+T_m). 
\end{equation}
Accordingly, each of the matrices in Eqn.~(\ref{EqMotionM}) can be expanded in terms of their Fourier series and expressed as
\begin{subequations}
\begin{eqnarray}
\bK^{(l)}(t) & = & \sum_{q = -\infty}^{\infty} e^{i q \omega_m t }\bK_{q}^{(l)}, \\
\bK(t) & = &\sum_{q = -\infty}^{\infty} e^{i q \omega_m t }\bK_{q}, \\
\bK^{(r)}(t) & = & \sum_{q = -\infty}^{\infty} e^{i q \omega_m t }\bK_{q}^{(r)}
\end{eqnarray}
\end{subequations}
where $\bK_{q}$, $\bK_{q}^{(l)}$ and $\bK_{q}^{(r)}$ are the corresponding matrix coefficients and $\omega_m = 2\pi/T_m$ is the 
frequency associated with the temporal modulation. 

\subsection{Spatio-temporal plane wave expansion for the estimation of dispersion}\label{AD}
\label{BlochBasedRevised}
The dispersion relations for the considered time-varying structure can be estimated by seeking for a plane wave solution with modulated amplitude, which is expressed as
\begin{equation}\label{BlochTheorem}
\bu_{n}(t)=\bm{a}(t) e^{i (n \kappa \lambda_m + \omega t)} ,
\end{equation}
where $\kappa$ is the wavenumber and 
$\bm{a}(t) = \bm{a}(t+T_m)$ is a periodic amplitude function in time. 
The frequencies in the 
displacement amplitude $\ba(t)$ depend on the stiffness modulation frequency $\omega_m = 2\pi/T_m$  and it can be expressed 
as a Fourier series in the form
\begin{equation}\label{ansatz}
\bm{a}(t) = \sum_{p = -\infty}^{\infty}\ba_p e^{ip\omega_m t}. 
\end{equation}
Based on the solution form expressed by Eqn.~\eqref{BlochTheorem}, the following relations hold
\begin{equation}\label{BlochTheorem1}
\bu_{n-1}(t)=e^{-i \mu }\bu_{n}(t) ,\qquad  \bu_{n+1}(t)=e^{i \mu}\bu_{n}(t),
\end{equation}
where $\mu= \kappa \lambda_m$ is the non-dimensional wavenumber. Substituting into Eqn.~\eqref{EqMotionM} gives
\begin{equation*}\label{EqMotionBloch1}
\bM \ddot{\bu}_{n}(t)
+ \left( \bK^{(l)}(t) e^{-i \mu } + \bK(t)  + \bK^{(r)}(t) e^{i \mu} \right) \bu_{n}(t) = \bzero,
\end{equation*}
which may be written as
\begin{equation}\label{EqMotionBloch2}
\bM \ddot{\bu}_{n}(t) + \bhK(\mu,t) \bu_{n}(t) = \bzero.
\end{equation}
with
\begin{equation}\label{EqMotionBloch3}
\bhK(\mu,t)  =  \bK^{(l)}(t) e^{-i \mu} + \bK(t)  + \bK^{(r)}(t) e^{i \mu}. 
\end{equation}

Next, substituting Eqn.~\eqref{ansatz} into Eqn.~\eqref{EqMotionBloch2} and performing harmonic balance, by collecting the terms 
with frequency $\omega+p\omega_m$ we reach
\begin{equation}\label{QEP}
-(\omega + p \omega_m)^2 \bM + \sum_{q=-\infty}^{\infty}  \bhK_q(\mu)  \bm{a}_{p-q}= \bzero.
\end{equation}
Choosing a truncation order $P$ for the displacement amplitude $\ba(t)$, i.e., $\ba_p = \bzero$ for $|p|>P$, Eqn.~\eqref{QEP} reduces to 
a system of $(2P+1)R$ equations, with $R$ being the number of degrees of freedom in a unit cell. This system of equations defines a
quadratic eigenvalue problem 
that can be solved in terms of frequency $\omega$ for assigned values of $\mu$. 
Specifically, $\mu$ spans the reciprocal lattice space, limited to $\mu \in [-\pi, \,\,\, +\pi]$ which defines the First Brillouin zone. 
It is here considered instructive to analyze the structure of the eigenvalue problem in Eqn.~\eqref{QEP}. By doing a change of variable 
$p\to p+r\omega_m$, the resulting expression may be written as
$-( (\omega+r\omega_m) + p \omega_m)^2 \bM + \sum_{q = -\infty}^{\infty} \bhK_q(\mu)  \bm{a}_{(p+r)-q} = \bzero$. Comparing it with 
Eqn.~\eqref{QEP}, we observe that if $(\omega,\ba_p)$ is an eigensolution of the infinite dimensional eigenvalue problem, then 
$(\omega+r\omega_m, \ba_{p+r})$ is also a solution. 

The solution of the following quadratic eigenvalue problem with truncated terms
\begin{equation*}
-(\omega + p \omega_m)^2 \bM + \sum_{q=-P+p}^{P+p}  \bhK_q(\mu)  \bm{a}_{p-q}= \bzero.
\end{equation*}
thus leads to  $R\times(2P+1)$ eigenvalues that are of the general form
\begin{equation}\label{EIG}
\lambda_{r,p} = \omega_r+ p \omega_m
\end{equation}
with $r=1,..,R$ and $p=-P,...,+P$. The associated eigvenvectors can be expressed as
\[
\bm \alpha_{r,p} = [\bm{a}_{-P}^{(p)}, ...\,\, ,\bm{a}_{0}^{(p)}, \,\, ..., \bm{a}_{+P}^{(p)} ]^T_r. 
\]
Equation~\eqref{EIG} shows how the eigenvalues are clustered into $R$ groups of $2P+1$ values centered at frequencies $\omega_r$ and separated by integer multiples of the modulation frequency $\omega_m$. Thus, the solution approach introduces the challenge of identifying 
the $R$ dispersion branches associated with the sought plane wave solution from the $R(2P+1)$ frequencies obtained from the eigenvalue problem. This challenge is here addressed through a procedure that identifies the plane wave branches based on a weighting 
factor corresponding to the magnitude of the eigenvector components associated with the fundamental plane wave term. 

To illustrate the rationale of this weighting procedure, we consider the plane wave solution imposed in Eqns.~\eqref{BlochTheorem} 
and \eqref{ansatz}. 
The $r$-th family of solutions ($2P+1$ solutions) associated with this plane wave may be expressed as
\begin{equation}\label{ansatz1}
\bm{u}^{(r)}_n(t) =e^{i n \mu} \sum_{p = -P}^{+P} \left(\sum_{q = -P}^{+P} \ba_q^{(p)}|_r e^{i (\omega_r+q\omega_m)t} \right) e^{i p\omega_m t}
\end{equation}
which can be re-organized to read
\begin{equation}\label{ansatz2}
\bm{u}^{(r)}_n(t) =e^{i (n \mu+\omega_r t)} \sum_{p = -P}^{+P} \left(\sum_{q = -P}^{+P} \ba_q^{(p)}|_r e^{i (p+q)\omega_m t} \right)
\end{equation}
Thus, the fundamental plane wave term has a magnitude that is identified for $p+q=0$. It is reasonable to expect that such fundamental is the leading term in the expansion such that 
\begin{equation}\label{ansatz3}
\bm{u}^{(r)}_n(t) \approx e^{i (n \mu+\omega_r t)} \sum_{p = -P}^{+P} \left(\sum_{q = -P}^{+P} \ba_q^{(p)}|_r \delta_{p+q,0}\right)
\end{equation}
where $\delta_{p+q,0}$ is the Kronecker delta, i.e. $\delta_{i,j}=1$ if $i=j$, or $\delta_{i,j}=0$ if $i\neq j$. According to the above equation, the dispersion branch corresponding to the fundamental plane wave can be effectively obtained weighting each branch by the magnitude of the fundamental component and applying a thresholding value which filters the branches, thereby avoiding their plotting if the associated eigenvector magnitude is low compared to that of the fundamental. The effectiveness of this procedure in tracking the branch corresponding to the plane wave of interest is illustrated through the examples presented in the next section. 


\subsection{Spectral energy density method}\label{SED}
We now briefly describe the spectral energy density method for obtaining the dispersion diagrams as used by other studies~\cite{swinteck2015bulk}. In this method, displacements are obtained by numerical integration. Full numerical simulations of the system velocity field under a range of excitation frequencies are needed, and the results are used to construct the dispersion diagrams. 

Consider a finite lattice having $R$ degrees of freedom per unit cell and a total length of $N$ unit cells, i.e. a total of $RxN$ degrees of freedom. For the algorithm, first the velocity field of each simulation $\dot{\bu}(n,t)$ is projected into the orthogonal Fourier basis $e^{i(\mu n - \omega t)}$
\begin{equation}\label{Eq.Fu}
\dot{\bu}(n,t)  = \sum_{\omega} \sum_{\mu} \bc^{(\omega,\mu)} e^{i(\mu n - \omega t)}. 
\end{equation}
\begin{equation}
\bc^{(\omega,\mu)}=\left\langle \dot{\bu}(n,t),e^{i(\mu n - \omega t)}\right\rangle =\dfrac{1}{TN}\int_0^T \sum_{n = 1}^N \dot{\bu}(n,t) e^{i(\mu n -\omega t)} dt.
\end{equation}
where $T$ is the total integration time. The spectral energy density is a measure of the energy content of the each plane wave with specific frequency $\omega$ and wavenumber $\mu$, by considering the kinetic energy of the masses, $ (1/2)m\lVert \bc^{(\omega,\mu)}\rVert^2$. Given Eqn.~\eqref{Eq.Fu}, the SED $\Phi(\omega,\mu)$ is obtained as 
\begin{equation}\label{Eq.SED.2}
\Phi(\omega,\mu) = \sum_{r = 1}^R \dfrac{1}{2} m \left( {c_r^{(\omega,\mu)}}\right) ^{2} = 
\dfrac{m}{2(TN)^2}\sum_{r=1}^R \left| \int_0^T \sum_{n = 1}^N \dot{u}_r(n,t) e^{i(\mu n -\omega t)} dt \right|^2 
\end{equation}
where $r$ is the position of the particle within the unit cell. We performed the time integration numerically using the trapezoidal rule
\begin{equation}\label{Eq.SED}
\Phi(\omega,\mu) = \dfrac{m}{2(TN)^2}\sum_{r=1}^R \left| \Delta t \sum_{j = 0}^{T/\Delta t} \sum_{n = 1}^N \dot{u}_r(n,j\Delta t) e^{i(\mu n -\omega j\Delta t)} \right|^2 
\end{equation}
where $\Delta t$ is the grid spacing of the numerical integration.

Last, Eqn.~\eqref{Eq.SED} is evaluated for each $(\omega,\kappa)$ pair and the result is a contour plot. The dispersion diagrams are given by the high contour values of $\Phi$, indicating the allowance of plane waves with that specified frequency/wavenumber pair.


\section{Verification of the proposed method: Dispersion analysis of spring-mass lattices}\label{Sec.3}
The procedure explained in Section \ref{TheorySection} is illustrated for the case of the spring-mass lattice shown in Fig. \ref{fig:DeymierChainSchematics_JustModulation}. The masses of the lattice are considered constant in time, and are all equal to $m$. The masses are connected to their adjacent neighbors by springs of time-modulated constant $k_r(t) = k_0+ k_m \Phi_r(t) $, and to the ground by springs that may be also time-modulated, \emph{i.e.} $k_{g,r}(t) = k_{g_0}+k_{g_m}\Phi_{g,r}(t)$. According to the derivations in the previous section, the procedure considers periodic time modulations, so that the two functions $\Phi(t), \Phi_m(t)$ are both periodic of period $T_m$, \emph{i.e.}  $\Phi_r(t)=\Phi_r(t+T_m), \Phi_{g,r}(t)=\Phi_{g,r}(t+T_m)$.

The equation of motion for the $r$-th mass is 
\begin{equation}\label{EqMotion}
m \ddot{u}_{r} + k_{r-1}(t) (u_{r} - u_{r-1})  + k_{r}(t)( u_{r} - u_{r+1} ) + k_{g,r}(t) u_{r} = 0.
\end{equation}

A characteristic frequency $\omega_0^2=k_0/m$ is introduced, so that Eqn.~\eqref{EqMotion} can be rewritten in non-dimensional form as follows
\begin{equation}\label{EqMotion_nd}
\ddot{u}_{r} + \omega_0^2 (1+\beta_m \Phi_{r-1}(t)) (u_{r} - u_{r-1})  + \omega_0^2 (1+\beta_m \Phi_{r}(t))( u_{r} - u_{r+1} ) + \omega_0^2 \gamma_g (1+\beta_{g,m} \Phi_{g,r}(t)) u_{r} = 0.
\end{equation}
where
\[
\beta_m=\frac{k_m}{k_0}, \,\, \beta_{g,m}=\frac{k_{g,m}}{k_g}, \,\, \gamma_{g}=\frac{k_{g}}{k_0}
\]

We note that  the above expression corresponds to 
a system of Mathieu-Hill equations, whose stability with respect to the modulation parameters has been extensively investigated~\cite{bolotin1962dynamic}.  
The method presented here applies when the parameters lie in the stable regions. In our examples, the modulation properties are small compared to  the constant order terms and we assume that solutions lie within the stable regions. 

Three examples of stiffness modulated systems are studied. 
In the first two examples, we analyze the dispersion of one-dimensional lattices having a harmonic modulation of their properties in space and time. 
In the last example, we study a spring-mass chain with a traveling square-wave modulation, thereby 
demonstrating the applicability of the method when there are multiple modulation frequencies.

\subsection{Harmonic stiffness modulation}\label{Sec.3.1}

\begin{figure}[hbtp]
	\centering
	\includegraphics[width=0.9\textwidth]{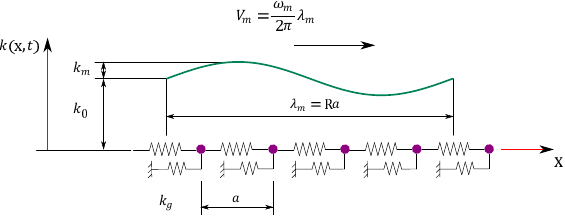}
	\caption{Schematic of a chain of resonators with harmonic modulation of the first-neighbor interaction springs. The modulation pattern travels with velocity $V_m =\omega_m \lambda_m/2\pi$.}
	\label{fig:DeymierChainSchematics_JustModulation}
\end{figure}

A first example considers a stiffness modulation that is imposed along the length of the chain with spatial wavelength $\lambda_m=Ra$, where $a$ is the distance between neighboring masses, while $R$ denotes the number of masses in a spatial modulation wavelength $\lambda_m$.  
Figure~\ref{fig:DeymierChainSchematics_JustModulation} displays a schematic of the chain along with the spring stiffness modulation. 

We first consider a periodic modulation of the springs between adjacent masses, so that the stiffness of the $r$-th spring is
\begin{equation}
k_r(t)=k_0 (1+\beta_m \cos\left(\omega_m t+\phi_r \right)), 
\end{equation}
where $\phi_r = 2 \pi r/R$ is the spatial phase shift related to the position $r$ of the considered spring. Also, the ground stiffnesses are considered constant in time, therefore in Eqn.~\eqref{EqMotion_nd} $\beta_{g,m}=0$. A unit cell is identified that contains $R$ degrees of freedom, whose behavior is governed by a system of equations of the kind of Eqn.~\eqref{EqMotionM}, subsequently transformed into Eqn.~\eqref{QEP}, with $\bM = \bI$, where $\bI \in \mathbb{R}^{R \times R}$ is the identity matrix, while
\begin{equation}
\begin{split}
\bhK_q (\mu) = & \gamma_g \omega_0^2  \bI \delta_{q,0} + \omega_0^2
\begin{bmatrix}
2 & -1 & 0 & 0 & \dots  & -e^{-i \mu} \\
-1 & 2 & -1 & 0 & \dots  & 0 \\
0 & -1 & 2 & -1 & \dots  & 0 \\
\vdots & \vdots & \vdots & \vdots & \ddots & \vdots \\
-e^{i \mu} & 0 & 0 & 0 & \dots  &  2
\end{bmatrix} \delta_{q,0} \\
+ &\omega_0^2 \frac{  \beta_m }{2}
\begin{bmatrix}
e^{iq\phi_R}+e^{q i\phi_1} & -e^{iq\phi_1} & 0 & \dots  &  -e^{ i q\phi_R}e^{-i\mu} \\
-e^{iq\phi_1} & e^{iq\phi_1}+e^{iq\phi_2} & -e^{iq\phi_2} & \dots  & 0 \\
0 & -e^{iq\phi_2} & e^{iq\phi_2}+e^{iq\phi_3} & \dots  & 0 \\
\vdots & \vdots & \vdots  & \ddots & \vdots \\
-e^{ i q\phi_R}e^{i\mu} & 0 & 0 & \dots  &  e^{iq\phi_{R-1}}+e^{q i\phi_R}
\end{bmatrix} \delta_{q,\pm 1}
\end{split}
\end{equation}
%

We study a chain with modulation wavelength $\lambda_m =3a$ ($R = 3$ masses in each unit cell) 
and analyze its dispersion properties estimated from the solution of Eqn.~\eqref{QEP}. Given the considered form of modulation, which is harmonic, the solution is conducted by considering $P =1$ terms, which is found to lead to results that are in agreement with those estimated through numerical simulations and for higher order expansions. Results are presented for the following set of parameters ${\Omega}_m={\omega_m}/{\omega_0}=0.2$, $\beta_m=0.15$, $\gamma_g=1$.

\begin{figure}[hbtp]
	\centering       
	\begin{subfigure}[b]{0.45\textwidth}
		\includegraphics[width=\textwidth]{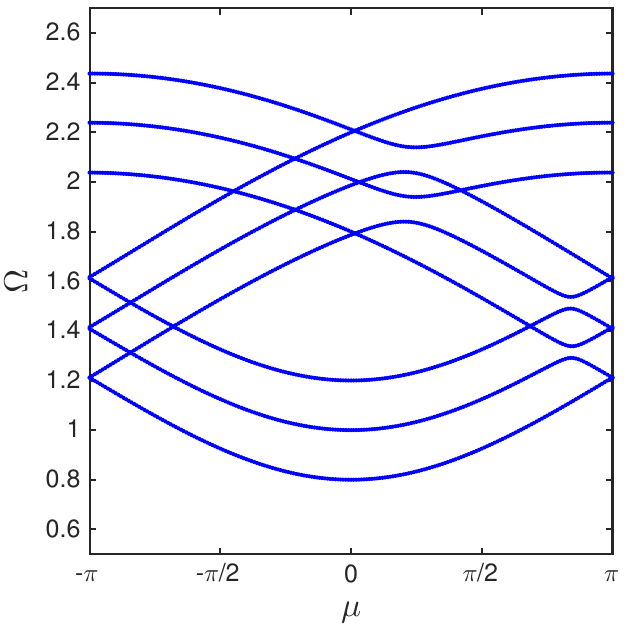}
		\caption{}
		\label{Fig.Deymier.BBDD.a}
	\end{subfigure}               \\
	\begin{subfigure}[b]{0.45\textwidth}
		\includegraphics[width=\textwidth]{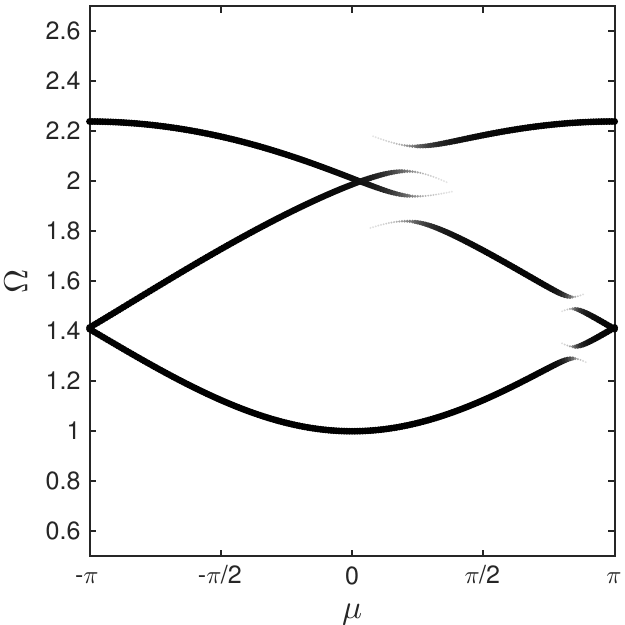}
		\caption{}
		\label{Fig.Deymier.BBDD.b}
	\end{subfigure}               
	\begin{subfigure}[b]{0.45\textwidth}
		\includegraphics[width=\textwidth]{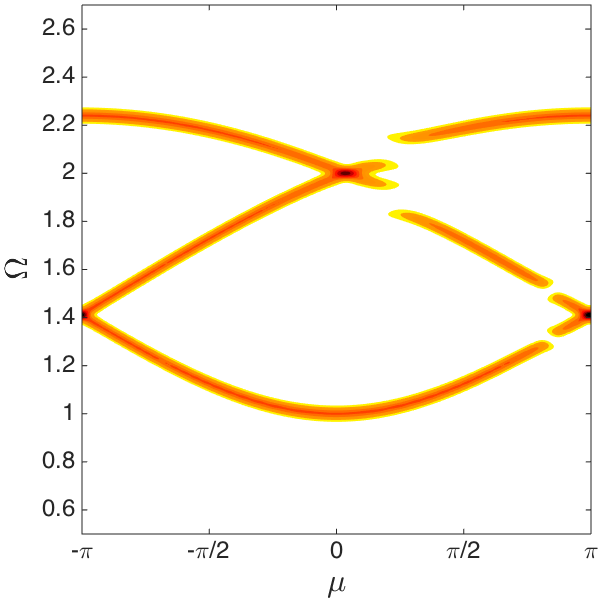}
		\caption{}
		\label{Fig.Deymier.BBDD.c}
	\end{subfigure}
	\caption{Branches resulting from the solution of the eigenvalue problem in Eqn.~\eqref{QEP} for $R=3$ and assigned $\mu \in [-\pi,\,\, +\pi]$ (a). Fundamental branch obtained through weighting and thresholding process (b) illustrating the presence of asymmetric bandgaps as a result of the harmonic spatio-temporal modulation of the stiffness constants. Dispersion diagram obtained with the SED method shows excellent agreement with the predicted fundamental dispersion branches (c).}
	\label{Fig.Deymier.BBDD}
\end{figure}

The solution is conducted by imposing $\mu \in [-\pi,\,\,\pi]$ which provides $R\times (2P+1)=9$ frequency branches. These branches are plotted in terms of the non-dimensional frequency $\Omega=\omega / \omega_0$ in Fig.~\ref{Fig.Deymier.BBDD.a}. Figure~\ref{Fig.Deymier.BBDD.a} displays $3$ families of curves that appear parallel and separated by $\Omega_m$. These branches cross for certain wavenumber values and fold at the boundaries of the first Brillouin zone. 
The dispersion diagram corresponds to the branch associated with the fundamental plane wave component among the $3$ families of curves. 
To identify this branch, the filtering method described in the previous section is applied, whereby the plotting of each curve is based on a weighting parameter defined by the amplitude of the eigenvector corresponding to the fundamental and a thresholding procedure. The result of this process is shown in Fig.~\ref{Fig.Deymier.BBDD.b}, which shows only the single fundamental branch. As expected the lattice is characterized by a high-pass cut-off at $\Omega =1$ for $\mu=0$ which is defined by the ground spring stiffness, and by a low-pass cut off for $\mu=\pm \pi$. Of significance is the asymmetry about the $\mu=0$ axis, which is generated by the modulation. This leads to asymmetric bandgaps seen in Fig.~\ref{Fig.Deymier.BBDD.b}, which result in the ability of the lattice to support one directional wave motion. For frequencies within one of these bandgaps, frequency is only defined for either a positive or negative wavenumber value which corresponds to waves propagating in the forward or backward direction only. For example, for $\Omega \in [2.05,2.15]$ the band diagram displays only a branch with negative slope, i.e., negative group velocity and thus excitation of the lattice within this frequency range will result in a wave propagating only in the $-x$ direction. Thus, lattices with periodic time-varying properties are characterized by non-reciprocal behavior for frequencies belonging to non-symmetric bandgaps. 

\begin{figure}[hbtp]
	\centering               
	\begin{subfigure}[b]{0.45\textwidth}
		\includegraphics[width=\textwidth]{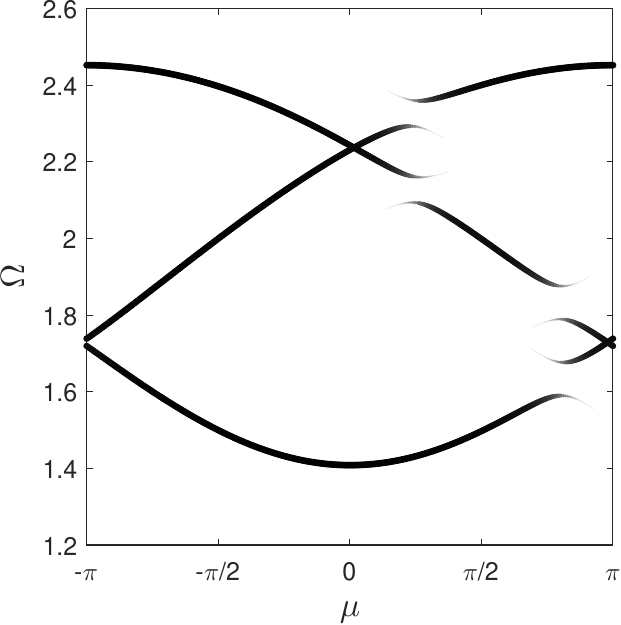}
		\caption{}
		\label{Fig.BDHG}
	\end{subfigure}\\
	\begin{subfigure}[b]{0.45\textwidth}
		\includegraphics[width=\textwidth]{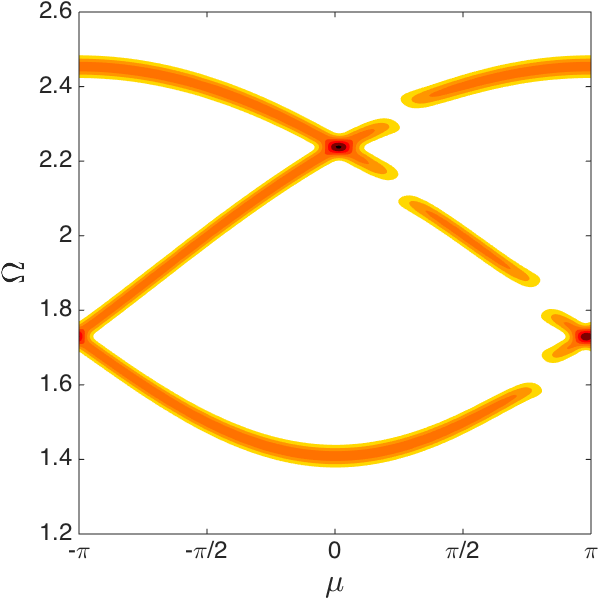}
		\caption{}
		\label{Fig.BDHG.SED}
	\end{subfigure}                      
	\caption{Dispersion diagrams for spring-mass lattice with modulated ground springs: fundamental branch evaluated through the procedure presented in this paper (a), and 
	results from the application of the SED method (b).}
	\label{Fig.BDHG.Comparison}
\end{figure}

We now compare the dispersion curves obtained using our Bloch based procedure with the dispersion data extracted from explicit numerical simulations using the SED method. 
The SED results are obtained by computing the transient response of the system to harmonic excitation at frequency $\Omega$, which is imposed as a prescribed displacement on one of the masses. The dispersion diagram is obtained through the excitation over a range of frequencies 
with each frequency 
excited individually. The velocity field obtained in each simulation is then used to compute the spectral energy density $\Phi(\Omega,\mu)$, which is shown as a contour plot to trace the dispersion diagram. In our calculation, we considered a closed system consisting of $70$ unit cells, whereby Born-Karman boundary conditions are enforced to avoid edge reflections~\cite{Polyzos2012}.
The results, displayed in Fig.~\ref{Fig.BDHG.Comparison}, show the excellent agreement between the SED method and the procedure presented in this paper.

The next example considers the modulation of the ground stiffness, which is expressed as
\begin{equation}\label{ground_stiffness}
k_{g,r}(t)=k_{g,0} (1+\beta_{g,m} \cos\left(\omega_m t+\phi_r \right)), 
\end{equation}
while the inter-mass stiffnesses are kept constant in time, \emph{i.e.} $\beta_{m}=0$. The corresponding stiffness coefficients are given by
\begin{eqnarray}
\nonumber \bhK_q (\mu) & = & \omega_0^2
\begin{bmatrix}
	 2 & -1 & 0 & 0 & \dots  & -e^{-i \mu} \\
	-1 & 2 & -1 & 0 & \dots  & 0 \\
	0 & -1 & 2 & -1 & \dots  & 0 \\
	\vdots & \vdots & \vdots & \vdots & \ddots & \vdots \\
	 -e^{i \mu} & 0 & 0 & 0 & \dots  &  2
\end{bmatrix} \delta_{q,0} + \gamma_g \omega_0^2  \bI \delta_{q,0} \\
& + & \gamma_g \omega_0^2 \frac{  \beta_{g,m}}{2}
\begin{bmatrix}
e^{iq\phi_1} & 0 & 0 & 0 & \dots  & 0 \\
0 & e^{iq\phi_2} & 0 & 0 & \dots  & 0 \\
0 & 0 & e^{iq\phi_3} & 0 & \dots  & 0 \\
\vdots & \vdots & \vdots & \vdots & \ddots & \vdots \\
0 & 0 & 0 & 0 & \dots  & e^{iq\phi_R}
\end{bmatrix} \delta_{q,\pm 1}
\end{eqnarray}
The considered lattice consists of unit cells with $R=3$ masses. Results are obtained for  ${\Omega}_m=0.2$, $\beta_{g,m}=0.15$, $\gamma_g=2$. For this set of parameters, the obtained dispersion branches are shown in Fig.~\ref{Fig.BDHG}, while the results obtained with the SED method are presented in Fig.~\ref{Fig.BDHG.SED}.  The results, displayed in Fig.~\ref{Fig.BDHG.Comparison}, show the excellent agreement between the SED method and the procedure presented in this paper.

\subsection{Square stiffness modulation}
\begin{figure}
	\centering
	\includegraphics[width=0.9\textwidth]{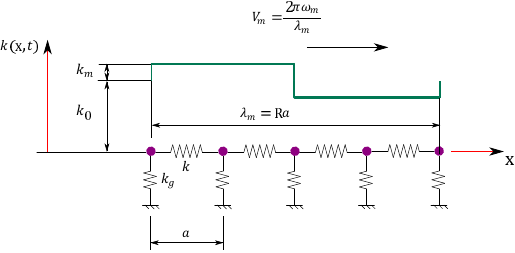}
	\caption{Schematic of a chain of resonators with a traveling square wave stiffness modulation.}
	\label{fig:DeymierChainSchematics_JustModulationSquare}
\end{figure}

Having demonstrated the effectiveness of our procedure in predicting dispersion diagrams for chains with harmonic modulation of stiffness, we now present a more complex example where the 
modulation is a traveling square wave and is thus composed of multiple harmonic components. The case of stiffness modulation imposed on the inter-mass springs is investigated for illustration purposes. The spring constant of the $s$-th spring is expressed as
\begin{equation}
k_r(t)=k_0 (1+\beta_m \big\{2H\big[\cos\left(\omega_m t+\phi_s \right)\big]-1\big\}), 
\end{equation}
where $H$ is the Heaviside function. A schematic of the modulated spring-mass chain is shown in Fig.~\ref{fig:DeymierChainSchematics_JustModulationSquare}.
Similar to the harmonic modulation, the stiffness modulation square wave travels with phase velocity $V_m=\omega_m \lambda_m/2 \pi$. The stiffness coefficients in Eqn.~\eqref{QEP} are now given by
\begin{eqnarray}
\nonumber \bhK_q (\mu) & = & \gamma_g \omega_0^2  \bI \delta_{q,0} + \omega_0^2
\begin{bmatrix}
	 2 & -1 & 0 & 0 & \dots  & -e^{-i \mu} \\
	-1 & 2 & -1 & 0 & \dots  & 0 \\
	0 & -1 & 2 & -1 & \dots  & 0 \\
	\vdots & \vdots & \vdots & \vdots & \ddots & \vdots \\
	 -e^{i \mu} & 0 & 0 & 0 & \dots  &  2
\end{bmatrix} \delta_{q,0} \\
& + &\omega_0^2 \beta_m 
\begin{bmatrix}
e^{iq\phi_R}+e^{iq\phi_1} & -e^{iq\phi_1} & 0 & \dots  &  e^{iq\phi_R}e^{-i \mu} \\
-e^{iq\phi_1} & e^{iq\phi_1}+e^{iq\phi_2} & -e^{iq\phi_2} & \dots  & 0 \\
0 & -e^{iq\phi_2} & e^{iq\phi_2}+e^{iq\phi_3} & \dots  & 0 \\
\vdots & \vdots & \vdots  & \ddots & \vdots \\
 -e^{iq\phi_R}e^{i \mu} & 0 & 0 & \dots  &  e^{iq\phi_{R-1}}+e^{iq\phi_Rß}
\end{bmatrix} c_q
\end{eqnarray}
where
\begin{equation}\label{squareModK}
c_q=\begin{cases} \frac{  2 }{|q| \pi},  & \text{for   } q\,\text{odd}\\
0, & \text{for   } q\,\text{even}
\end{cases}
\end{equation}

\begin{figure}
	\centering               
	\begin{subfigure}[b]{0.45\textwidth}
		\includegraphics[width=\textwidth]{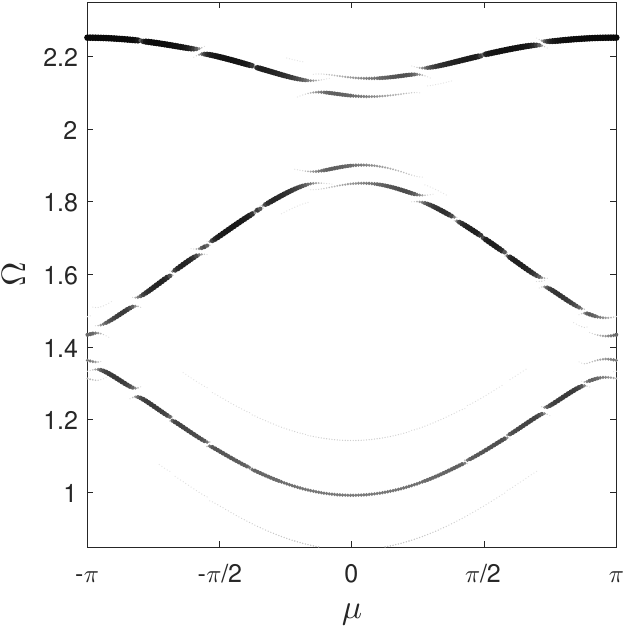}
		\caption{}
		\label{Fig.SqWave.a}
	\end{subfigure}\\
	\begin{subfigure}[b]{0.45\textwidth}
		\includegraphics[width=\textwidth]{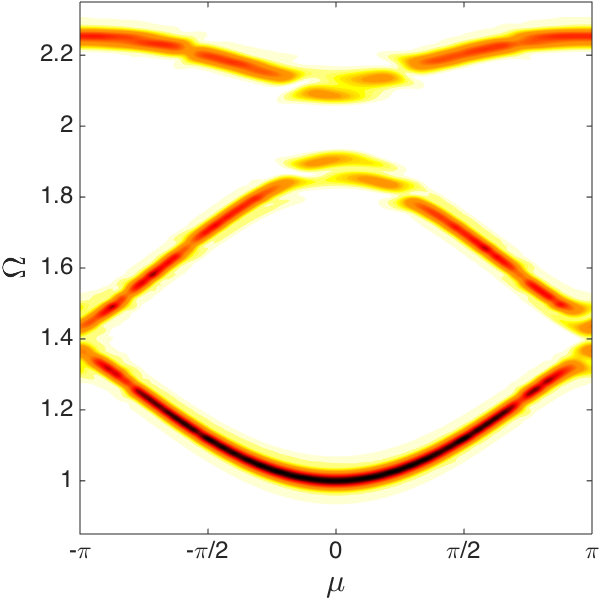}
		\caption{}
		\label{Fig.SqWave.b}
	\end{subfigure}                      
	\caption{Dispersion diagrams of the square-wave stiffness modulated lattice: dispersion diagram obtained with $2P+1=31$ coefficients (a), and SED method results (b).}
	\label{Fig.SqWave}
\end{figure}

The dispersion diagram of a lattice with modulation wavelength $\lambda_m=3a$ ($R=3$) 
and stiffness parameters ${\Omega}_m=0.25$, $\beta_{m}=0.05$, $\gamma_g=1$ is shown in Fig.~\ref{Fig.SqWave.a}, while the SED calculation is presented in Fig.~\ref{Fig.SqWave.b}. The number of terms used in the series expressing the stiffness modulation of the structure in Eqn.~\eqref{squareModK} is $2P+1=31$ ($P=15$). Since the square wave is a superposition of multiple frequencies, a larger number of terms than in the harmonic stiffness modulation case are required to obtain an accurate dispersion diagram for this system. The convergence rate of the Fourier series with respect to the number of terms is slow due to the Gibbs phenomenon when representing the square wave function in the Fourier basis. 

The square-wave modulated system also exhibits asymmetric dispersion behavior illustrating that the system violates time reversal symmetry. Note that the dispersion diagram is different from the harmonically modulated chain and displays several gaps in each branch instead of one due to the presence of multiple frequencies in the modulation.
A comparison of Figs.~\ref{Fig.SqWave.a} and~\ref{Fig.SqWave.b} shows that the proposed method predicts the gaps associated with stiffness modulation accurately. 
Finally, we note that the weighting process shows some traces of branches at low frequencies and low wavenumbers. There are two 
parallel branches to the actual dispersion curve and these arise due to the corresponding components of the eigenvector having a non-negligible amplitude. These additional branches may filtered out through a more aggressive threshold, which would need to be adapted in the case of multiple component modulations. Despite these minor traces, good agreement in Fig. \ref{Fig.SqWave} indicates that the Bloch-based analysis accurately predicts the dispersion diagrams where the modulation involves multiple frequencies. 


\section{Non-reciprocal behavior of time-modulated lattices}\label{Sec.4}
Upon observation of the time reversal symmetry breaking properties of modulated lattices, we now illustrate their implications on wave motion and dynamic response of both infinite and finite lattices. The discussion first illustrates the effects of the modulation parameters on the observed asymmetric gaps and relates them to the bandgaps observed in space-periodic, time-invariant systems. Next, numerical simulation of the transient response of modulated lattices illustrate the one-directional wave properties of this class of lattices, which are predicted by the analysis of dispersion. Finally, the study of a finite lattice shows how frequency modulation leads to the non-reciprocal steady-state response, which can be also predicted from observations made on the dispersion diagrams.

\subsection{Analysis of symmetry and asymmetry of the dispersion diagram}
\begin{figure}[hbtp]
	\centering               
	\begin{subfigure}[b]{0.45\textwidth}
		\includegraphics[width=\textwidth]{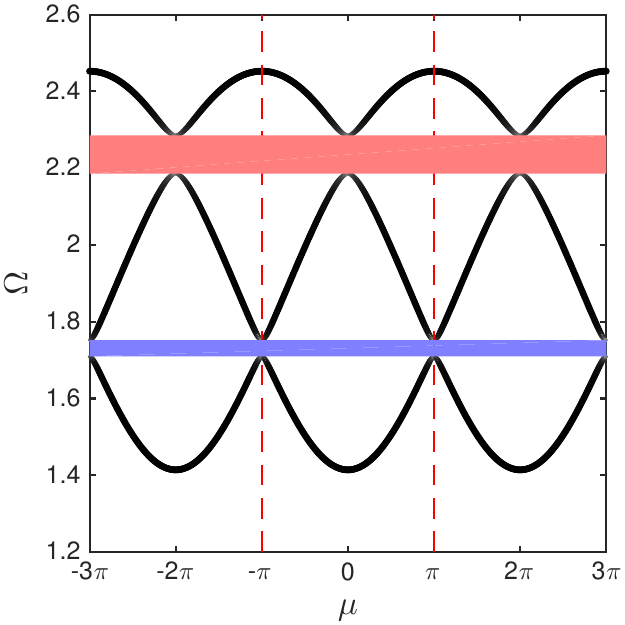}
		\caption{}
		\label{Fig.PeppeComplete.BBDD.a}
	\end{subfigure}\\
	\begin{subfigure}[b]{0.45\textwidth}
		\includegraphics[width=\textwidth]{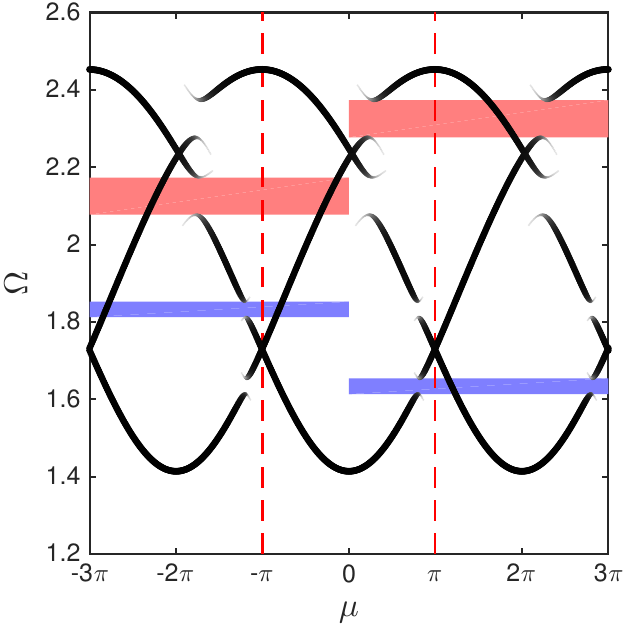}
		\caption{}
		\label{Fig.PeppeComplete.BBDD.b}
	\end{subfigure}                      
	\caption{Dispersion diagrams for (a) space modulated and (b) space-time modulated ($\lambda_m=3a$) systems having $3$ identical masses in a unit cell. 
	Space-time modulation produces asymmetric shift in the bandgaps. }
	\label{Fig.PeppeComplete.BBDD}
\end{figure}

The analysis of symmetries and their lack of thereof in the dispersion diagram due to spatio-temporal modulations is conducted with reference to the spring-mass lattice with harmonic ground stiffness modulation, whose parameters and stiffness properties are described in Section \ref{Sec.3} and Eqn.~\eqref{ground_stiffness}. For reference purposes we first consider the case where only spatial modulation exists, \emph{i.e.} when $\Omega_m = 0$. This leads to a time-invariant periodic system, which is known to feature a series of bandgaps whose frequency location and width are defined by the amplitude and period of the modulation. In this case, the period is $\lambda_m = 3 a$, while the modulation amplitude is $\beta_m=0.2$. The dispersion diagram for this lattice, shown in Fig.~\ref{Fig.PeppeComplete.BBDD.a}, shows the presence of two bandgaps, shaded in red and blue for visualization purposes, which are commonly expected to occur for a system with periodic spatial variations in mechanical properties. The diagrams are plotted over an extended wavenumber range, $\mu \in [-3\pi,+3\pi]$, to illustrate its periodicity as well as to highlight its symmetry with respect to the $\mu=0$ axis. 
In the presence of a modulation that occurs both in space and in time, defined in this case by a modulation frequency $\Omega_m=0.2$, the symmetry of the dispersion diagrams is broken as shown in Figure~\ref{Fig.PeppeComplete.BBDD.b}. Two asymmetric gaps occur at different frequencies for waves traveling to the left or to the right. The difference in the central frequency of the gaps, $\Delta\Omega= \Omega_m= 0.2$, is thus exactly equal to the value of the modulation frequency. This statement, which was proven analytically in~\cite{Trainiti2016} 
for continouus media, can be explained intuitively by observing that the impedance mismatch encountered by the propagating wave travels at velocity $v_m=\omega_m/\kappa_m$, so the rate at which the same wave perceives it is different when the wave travels to the left or to the right, by a difference that is the velocity of the modulation wave. Another perspective to this observation is to 
apply a change in variable $\bar x=x-(\omega_m/\kappa_m)t$ to the space-time modulated lattice, 
which recovers a space-only modulated lattice, but in the new coordinate system where the frequency of any traveling wave is shifted by a quantity $\omega_m$. One interesting observation is that the impedance mismatch remains the same, therefore the gaps have the same width, but occur at shifted frequencies. Similar observations are reported 
by Cassedy and Oliner~\cite{cassedy1963dispersion} in continuous systems subjected to space-time periodic modulation, where a proportional relation between the difference in frequency in the gaps ($\Delta \Omega$) and the phase velocity of the modulation wave $v_m$ was also derived.

\subsection{One-directional wave motion in modulated lattices}
\begin{figure}[hbtp]
	\centering
	\includegraphics[width=0.45\textwidth]{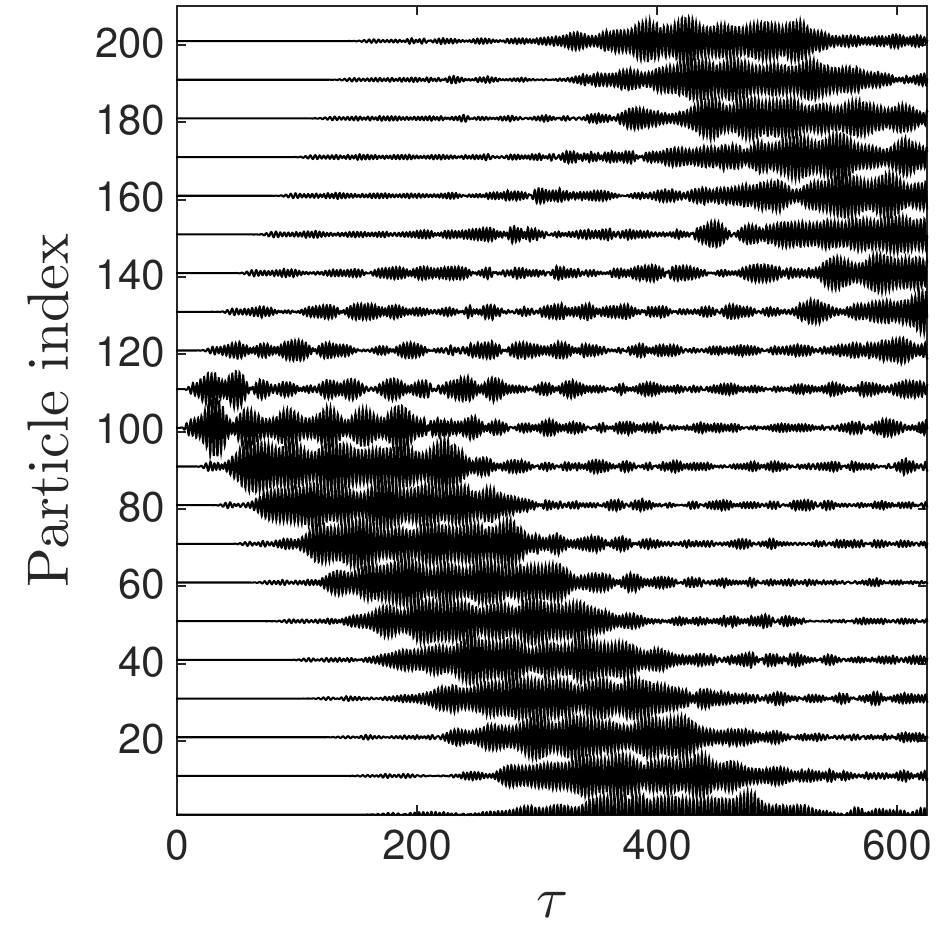}
	\caption{Contours of displacement magnitude for transient response to harmonic excitation of a stiffness modulated lattice. 
The waves propagate primarily in $-x$ direction, as predicted by the dispersion analysis.}
	\label{Fig.Deymier.NonReciprocal}
\end{figure}

The implications of asymmetric dispersion properties in terms of wave motion can be illustrated through numerical simulations conducted on lattices with the considered spatio-temporal modulations. 
These lattices violate mechanical reciprocity by allowing one-directional wave propagation. Consider a lattice with harmonic modulation of the first-neighbor interaction springs as discussed in Sec.~\ref{Sec.3.1}, having the properties $\Omega_m=0.2$, $\beta_m=0.15$, $\gamma_g=1$. The dispersion diagram of this lattice  exhibits an asymmetric bandgap in the branch with positive slope for $\Omega \in [2.05,2.15]$, see Fig.~\ref{Fig.Deymier.BBDD.b}. At frequencies in this bandgap, only modes with negative group velocities are allowed, which means that the lattice allows propagation in the $-x$ direction. The response of the lattice is evaluated for transient excitation over one mass at a frequency inside this bandgap, $\Omega=2.1$. For the numerical simulations, a lattice with $70$ unit cells and a modulation period $\lambda_m = 3a$ is considered, which corresponds to a total of $210$ masses.
Reflections at the edges are avoided by applying Born-Karman boundary conditions. Figure~\ref{Fig.Deymier.NonReciprocal} displays the evolution of displacement along the chain against the normalized time $\tau = \omega_0 t$. In the simulations, the center mass is excited for a duration of $\omega_0 t = 200$. The total simulation time is $\omega_0 t = 600$.
For the integration, we used Verlet algorithm \cite{Verlet1967} with a time step $\Delta t = 2\times 10^{-8}$.  
The resulting displacement field is strongly asymmetric, as most of the energy of wave motion is supported by waves that travel along the $-x$ direction. 
Some waves with frequency different from $\omega$ are also excited and they travel in both directions with group velocities different from the primary wave. Their amplitude is small compared to the primary wave, as can be observed in Fig.~\ref{Fig.Deymier.NonReciprocal}. Note that the dispersion diagrams only predict the wavenumbers and group velocities of 
waves traveling with the excitation frequency $\omega$ and the transient response of the chain is indeed consistent with the dispersion analysis. 

\subsubsection{Steady state response of a finite modulated lattice: acoustic circulator}

\begin{figure}[hbtp]
	\centering
	\includegraphics[width=0.8\textwidth]{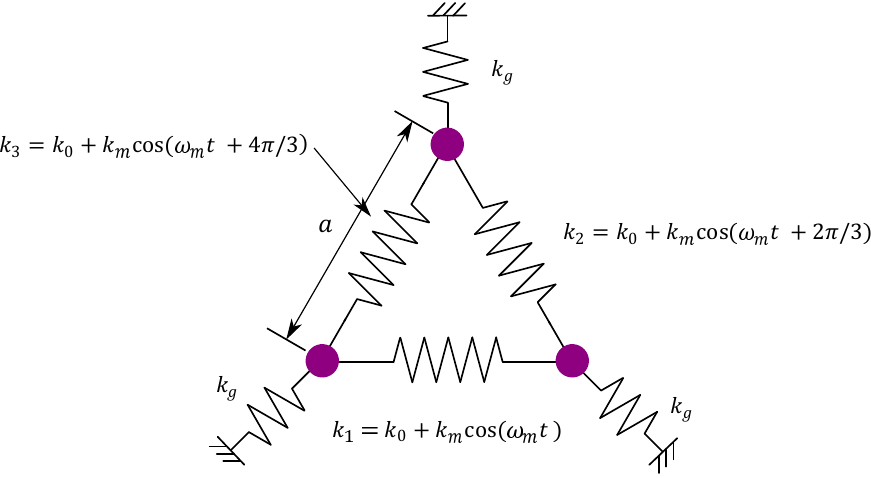}
	\caption{Schematic of a three mass system, with a stiffness modulation corresponding to a traveling wave applied on the springs.}
	\label{3massSchematic}
\end{figure}

Finally, we consider the free out-of-plane vibration response of a three mass system arranged in a circle as illustrated in Fig.~\ref{3massSchematic}. 
The masses are all identical and equal to $m$. Without loss of generality, the ground spring stiffness is set to $k_g=0$. 

The spring stiffness has two components: a constant part and a time varying modulation component. 
A traveling wave modulation of frequency $\omega_m$ and amplitude $k_m$ is applied to the springs connecting the masses and their 
stiffness is $k_r = k_0 + k_m\cos (\omega_m t + 2 \pi r /3)$. The governing equations of this system read $\bM \ddot{\bu} + \bhK \bu = \bzero$. They can be written in dimensionless form as $\ddot{\bu} + \bK \bu = \bzero$, where the stiffness matrix $\bK$ is 
\begin{equation}
\bK = {\omega_0}^2\begin{bmatrix} 2 & -1 & -1 \\ -1 & 2 & -1 \\ -1 & -1 & 2 \end{bmatrix} - {\omega_0}^2\dfrac{\beta_m}{2}\left( 
\begin{bmatrix} 1 & e^{i2\pi/3} & e^{i4\pi/3} \\ e^{i2\pi/3} & e^{i4\pi/3} & 1 \\ e^{i4\pi/3} & 1 & e^{i2\pi/3} \end{bmatrix}e^{i\omega_m t} + c.c\right),
\end{equation}
and $c.c$ stands for complex conjugate. To solve the system of equations, we follow a procedure similar to that of Alu and 
coworkers~\cite{estep2014magnetic}. The following change of variables is introduced to diagonalize the above system
\begin{equation}\label{transf}
\begin{bmatrix} u_1 \\ u_2 \\  u_3 \end{bmatrix}  = \dfrac{1}{\sqrt{3}} \begin{bmatrix}
1 & 1 & 1\\ 1 & e^{i2\pi/3} & e^{i4\pi/3} \\ 1 & e^{i4\pi/3} & e^{i2\pi/3} \end{bmatrix} 
\begin{bmatrix} a_1  \\ a_2 \\ a_3 \end{bmatrix}
\end{equation}
Note that we chose a basis that is not orthogonal because, as demonstrated below, the solution 
is expressed conveniently in the above basis in the presence of modulation. 
Using the transformation in Eqn.~\eqref{transf} leads to the following equation for $\ba = [a_1 \; a_2 \; a_3]^T$
\begin{equation}\label{govE_a}
\ddot{\ba} + {\omega_0}^2\left( \bK_0 + \dfrac{\beta_m}{2}\bK_1 \right) \ba = \bzero
\end{equation}
where 
\begin{equation}
\bK_0 = \begin{bmatrix} 0 & 0 & 0 \\ 0 & 3 & 0 \\ 0 & 0 & 3 \end{bmatrix}, \;\;\;
\bK_1 = \begin{bmatrix} 0 & 0 & 0 \\ 0 & 0 & 3 e^{ i(\omega_m t + 2\pi / 3) }\\ 0 & 3 e^{ -i(\omega_m t + 2\pi / 3) } & 0 \end{bmatrix}.
\end{equation}
Note that the $a_1$ mode becomes uncoupled from the $a_2$ and $a_3$ modes. To solve for these modes, we assume a solution of  the form 
\begin{equation*}
a_2 = A_2 e^{ i(\omega + \omega_m/2)t }, \;\;
a_3 = A_3 e^{ i(\omega - \omega_m/2)t }, 
\end{equation*}
where $\omega$ is an unknown frequency and $A_2, A_3$ are scalar constants. Substituting the above equation into the system of 
equations~\eqref{govE_a} results in a homogeneous system of equations and the condition for the existence of non-trivial solutions leads to  
\begin{equation}
\det \begin{pmatrix} 3 - (\Omega + \Omega_m/2)^2 & (3 \beta_m/2) e^{i2\pi/3} \\ (3 \beta_m/2) e^{-i2\pi/3} & 3 - (\Omega - \Omega_m/2)^2 
\end{pmatrix} = 0. 
\end{equation}
Solving the above system leads to the frequencies 
\begin{equation}\label{omegSoln}
\Omega_{1,2} = \left( \dfrac{\Omega_m^2}{4}  + 3 \pm \sqrt{3 \Omega_m^2 + \dfrac{9\beta_m^2}{4}} \right)^{1/2}. 
\end{equation}

Substituting values $\beta_m = 0.09$ and $\Omega_m = 0.5$ for the modulation parameters, we find that $A_2 \gg A_3$ for the mode $\Omega_1$ and 
$A_3 \gg A_2$ for the other mode $\Omega_2$. Thus transforming back to the original displacement basis using the inverse of the transform in Eqn.~\eqref{transf}, 
and considering only the term dominant between $A_2$ and $A_3$ in each of the modes, the solution is expressed as 
\begin{equation}
\bu = A_2 e^{i(\Omega_1 + \Omega_m/2)\tau} \begin{bmatrix}1 \\ e^{i2\pi/3} \\ e^{i4\pi/3} \end{bmatrix} 
			+ A_3 e^{i(\Omega_2 - \Omega_m/2)\tau} \begin{bmatrix}1 \\ e^{-i2\pi/3} \\ e^{-i4\pi/3} \end{bmatrix}. 
\end{equation}
Note that the frequencies split as a consequence of modulation from $\Omega= \sqrt{3}$ and the $a_2$ and $a_3$ 
modes are called clockwise and counter-clockwise propagating modes, respectively, based on the sign of the group velocity. 

\begin{figure}
	\centering               
	\begin{subfigure}{0.45\textwidth}
		\includegraphics[width=\textwidth]{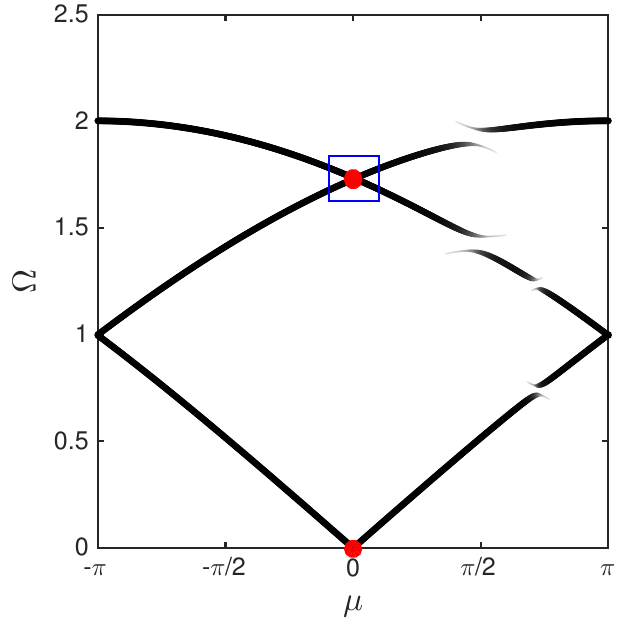}
		\caption{}
		\label{Steady_Fig1}
	\end{subfigure}
	\begin{subfigure}{0.45\textwidth}
		\includegraphics[width=\textwidth]{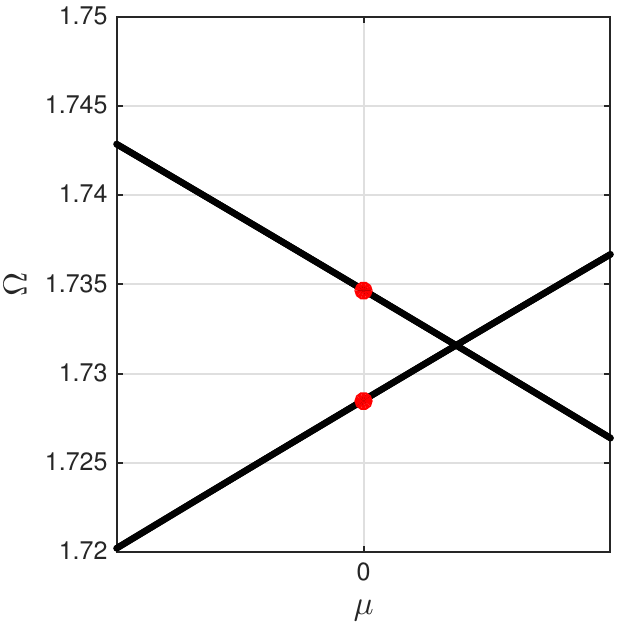}
		\caption{}
		\label{Steady_Fig2}
	\end{subfigure}                      
	\caption{(a) Dispersion diagram for a chain subjected to stiffness modulation showing loss of symmetry. (b) Zoomed in view of the dispersion diagram showing frequency splitting at $\mu = 0$. The left and right propagating waves have different frequencies and match with the frequencies obtained from steady state solution.}
	\label{Steady_Figs}
\end{figure}

We now interpret the above results using the dispersion analysis of an infinite chain of masses subjected to the same modulation. 
The normalized modulation stiffness $\beta_m$ and frequency $\Omega_m$ are set to $0.09$ and $0.5$, respectively. Figure~\ref{Steady_Fig1}
displays the dispersion diagram for a chain having a modulation wavelength $\lambda_m = 3a$. Similar to the previous case, there are 
$3$ branches in the dispersion diagram as a unit cell has three degrees of freedom. Also, the presence of modulation breaks the symmetry about 
the $\mu = 0$ axis, where $\mu=3 a \kappa$. Figure~\ref{Steady_Fig2} displays a zoomed-in view of the dispersion diagram where the two branches intersect close to $\mu = 0$. 
We observe that the branches do not intersect at $\mu = 0$ as a result of modulation. 

\begin{figure}[hbtp]
	\centering
	\includegraphics[width=0.5\textwidth]{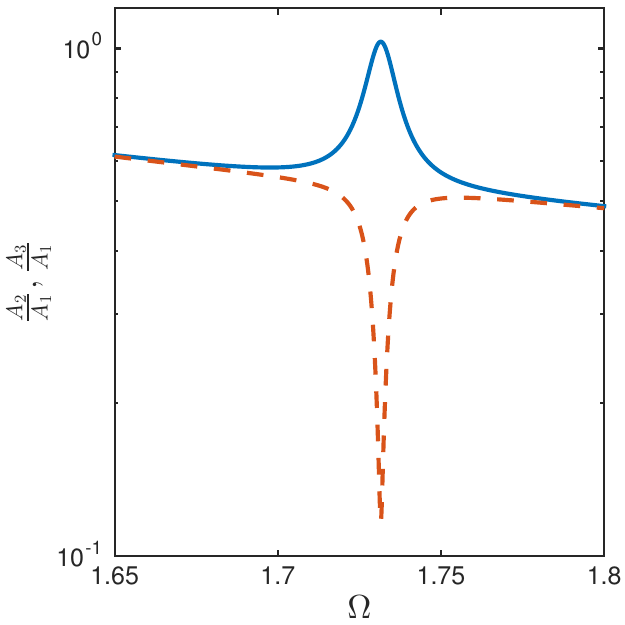}
	\caption{Frequency response function when particle $1$ is excited. Amplitudes ratio $A_2/A_1$, $A_3/A_1$ in a blue solid and a red dashed line respectively. At excitation frequency $\Omega=1.7315$ the amplitude of particle 3 response is much smaller than the other two.}
	\label{FRF}
\end{figure}

The steady state free vibration response corresponds to a constraint $u_i=u_{i+3}$ imposed on an infinite chain. This constraint is satisfied for wavenumbers 
$\mu$ which have $e^{i\mu} = 1$. This condition is equivalent to $\mu/2 \pi = n$, where $n$ is an integer and is equivalent to $\mu= 0$ in the 
First Brillouin zone. Figure~\ref{Steady_Figs} displays that there are 3 values of frequency $\Omega$ at wavenumber $\mu = 0$. One frequency is $\Omega= 0 $, while there are
two non-zero frequencies. In the absence of modulation, these frequencies would be $\Omega= \sqrt{3}$. Modulation results in a split of frequencies, and the two frequencies are derived in Eqn.~\eqref{omegSoln}. From the dispersion diagram, we infer that the wave with frequency 
$\Omega_2-\Omega_m/2=1.7285$ is a right traveling wave while the wave with frequency $\Omega_1+\Omega_m/2=1.7347$ is a left traveling wave. The dispersion analysis thus predicts
accurately the frequency splitting due to stiffness modulation in a 3-mass system. This frequency splitting is the basis for designing non-reciprocal acoustic circulators. When the 
system is excited at the frequency $\Omega=1.7315$, the combination of the $a_2$ and $a_3$ modes results in small ratio $A_3/A_2$, as demonstrated by Alu and
coworkers~\cite{estep2014magnetic}, see frequency response function in Fig. \ref{FRF} (damping coefficient was set to c=0.01).
This phenomena can be exploited to build a nonreciprocal mechanical $3$-port device. 

\section{Conclusions}\label{ConcSection}
The paper presents a Bloch-based method to study the dispersion properties of systems with periodic time-varying coefficients.
The method allows us to obtain the dispersion diagram of the lattice by solving a quadratic eigenvalue problem over the unit cell, whose length depends on the spatial modulation wavelength. The applicability of our method is demonstrated for both harmonic modulation and modulation having multiple frequencies, and it is validated with 
dispersion diagrams obtained from numerical simulations using SED method. 
Our method shows excellent agreement with the dispersion diagrams obtained from full numerical simulations for all the considered cases and predicts unidirectional wave propagation at 
certain frequencies. 
Potential future research directions include extending our formulation to higher dimensions and wave propagation in other physical systems, as well as designing lattices with specific 
nonreciprocal mechanical wave propagation characteristics. 

\section{Acknowledgements}
The authors are indebted to the {\it US Army Research Office} (Grant number W911NF1210460), the {\it US Air Force Office of Scientific Research} (Grant number FA9550-13-1-0122), the {\it University Carlos III de Madrid} and the {\it Ministerio de Ciencia e Innovaci\'{o}n de Espa\~{n}a} (Project DPI/2014-57989-P) for the financial support.

\bibliographystyle{unsrt}
\bibliography{papers}

\end{document}